\documentclass[journal]{IEEEtran}

\usepackage{cite}

\usepackage[mode=buildnew]{standalone}% requires -shell-escape
\usepackage{tikz}

\usepackage{amsmath}
\usepackage{amsfonts}
\usepackage{algorithmic}
 \usepackage{graphicx}
  \ifCLASSOPTIONcompsoc
 \usepackage[caption=false,font=normalsize,labelfont=sf,textfont=sf]{subfig}
 \else
 \usepackage[caption=false,font=footnotesize]{subfig}
 \fi
\usepackage{url}
\usepackage{booktabs}       % professional-quality tables
\newcommand{\includeCroppedPdf}[2][]{%
    \IfFileExists{./#2-crop.pdf}{}{%
        \immediate\write18{pdfcrop #2 #2-crop.pdf}}%
    \includegraphics[#1]{#2-crop.pdf}}

\renewcommand{\vec}[1]{\mathbf{#1}}

\begin{document}

\title{Place of Occurrence of COVID-19 Deaths in the UK: Modelling and Analysis}

\author{\IEEEauthorblockN{Spencer A. Thomas\IEEEauthorrefmark{1}%,~\IEEEmembership{Member,~IEEE} 
}
\IEEEauthorblockA{\IEEEauthorrefmark{1}National Physical Laboratory, Teddington, UK}
% <-this % stops an unwanted space
\thanks{Manuscript received XX XX, XXXX; revised XX XX, 2020. 
Corresponding author: S. A. Thomas (email: spencer.thomas@npl.co.uk).}}

% The paper headers
\markboth{Modelling COVID-19 Deaths in the UK}%
{Shell \MakeLowercase{\textit{et al.}}: Bare Demo of IEEEtran.cls for IEEE Transactions on Magnetics Journals}
% The only time the second header will appear is for the odd numbered pages
% after the title page when using the twoside option.

\IEEEtitleabstractindextext{%
\begin{abstract}
We analysed publicly available data on place of occurrence of COVID-19 deaths from national statistical agencies in the UK between March 9 2020 and February 28 2021. We introduce a modified Weibull model that describes the deaths due to COVID-19 at a national and place of occurrence level. We observe similar trends in the UK where deaths due to COVID-19 first peak in Homes, followed by Hospitals and Care Homes 1-2 weeks later in the first and second waves. This is in line with the infectious period of the disease, indicating a possible transmission vehicle between the settings. Our results show that the first wave is characterised by fast growth and a slow reduction after the peak in deaths due to COVID-19. The second and third waves have the converse property, with slow growth and a rapid decrease from the peak. This difference may result from behavioural changes in the population (social distancing, masks, etc). Finally, we introduce a double logistic model to describe the dynamic proportion of COVID-19 deaths occurring in each setting. This analysis reveals that the proportion of COVID-19 deaths occurring in Care Homes increases from the start of the pandemic and past the peak in total number of COVID-19 deaths in the first wave. After the catastrophic impact in the first wave, the proportion of COVID-19 deaths occurring in Care Homes gradually decreased from is maximum after the first wave indicating residence were better protected in the second and third waves compared to the first.

\end{abstract}

% Note that keywords are not normally used for peerreview papers.
}

% make the title area
\maketitle
\IEEEdisplaynontitleabstractindextext
\IEEEpeerreviewmaketitle

% % separate document in submission
% \section{Highlight}
% \begin{itemize}
%     \item Deaths due to COVID-19 in the UK follow a Weibull distribution
%     \item Deaths due to COVID-19 peaked first in Homes first followed by Hospitals and Care Homes 1-2 weeks later
%     \item The first wave was characterised by fast growth and slow reduction in deaths due to COVID-19
%     \item The second and third waves exhibited slow growth and rapid reduction in deaths due to COVID-19 
%     \item The proportion of COVID-19 deaths in Care Homes increased beyond the first wave peak. 
% \end{itemize}

\section{Introduction}
\IEEEPARstart{T}{he} coronavirus pandemic is one of the biggest public health issues ever experienced. The rapid spread of COVID-19, which is caused by Severe acute respiratory syndrome coronavirus 2 (SARS-CoV-2) \cite{lai2020}, has led to worldwide spread \cite{rodriguez-morales_going_2020} and significant interventions from governments around the world \cite{balmford_cross-country_2020}. In the UK, the government issued a national lockdown in March 2020 to stop the spread. Following an easing of national restrictions, experts are warning of a successive waves of COVID-19 in the UK. The first wave of COVID-19 in the UK had a devastating impact. The catastrophic effect in Care Homes across the UK was widely reported \cite{guardianDoubleDeath, guardianDoubleSecrect, HealthFoundation} due to a greater risk of outbreak \cite{guardianOutbreak}, excess death \cite{BBCexcessDeath} and a delay in reporting of the data \cite{iacobucci_covid-19-2020}. The impact on hospitals has been significant. With resources focused on COVID patients other healthcare needs, such as surgeries, cancer treatments, and routine appointments, have been deprioritised. The recent vaccines could help the fight against the pandemic, though regulatory and logistical barriers mean that this will not be a short term solution. As COVID-19 cases and deaths continues to rise in the UK, and with the emergence of new strains of the virus, understanding the trends in the disease on a population level can help national and regional policies to prevent further deaths. Analysis of population level data can lead to improvements for disease incidence estimates, prediction of infection prevalence, and decision making at a national scale \cite{Souty2014}.

A large number of studies have focused on using medical imaging data to diagnosing COVID-19 patients or classification of the severity of the disease \cite{abbas_classification_2020,sahlol_covid-19_2020,wu_severity_2020,qian_m3lung-sys_2020,meng_deep_2020,li_classification_2020,tabik_covidgr_2020}. These studies have aimed to better understand the markers to identify COVID-19 and alleviate the workload for medical professionals, though a recent review found that methods are currently not of clinical use due to methodological flaws or underlying biases \cite{roberts_common_2021}. Beyond imaging data, other studies analysing COVID-19 data have focused on infection or transmission of the disease \cite{ferretti_quantifying_2020}, comparing strategic response \cite{gibney_whose_2020}, or identifying risk factors \cite{lusignan_risk_2020}. Tracking and modelling of infection or mortality rates is challenging due to asymptomatic cases and testing accuracy \cite{grassly_comparison_2020,WHO_covid_mortality2020}. This was particularly difficult at the start of the pandemic where community testing was not available. %In contrast, there is more confidence in COVID-19 mortality data.\cite{WHO_covid_mortality2020} 
More recent studies have looked at the effects of COVID-19 and national lockdown on mental health \cite{lemanska_study_2021,wang_impact_2021, mansfield_indirect_2021}.
To date there have been few studies on how COVID-19 mortality varies in different settings within a nation. Analysis of the pandemic in different counties or regions have compared the evolution of the disease \cite{acter_evolution_2020} or government responses to the pandemic \cite{han_lessons_2020,balmford_cross-country_2020}. However, we have found no studies looking at different healthcare and domestic settings within a nation, or the trends in mortality due to COVID-19.

We aim to model the ``place of occurrence'' (Hospitals, Care Homes, Homes, etc) of COVID-19 deaths in the UK. We analyse data available from national statistical agencies throughout the pandemic. We consider both deaths due to COVID-19 (where COVID-19 is mentioned on the death certificate) and deaths from all causes.

\section{Methods}
\label{sec:methods}

In this work we focus specifically on the place of occurrence of death for both COVID-19 and all deaths from national statistical agencies. COVID-19 related deaths include any death where Coronavirus or COVID-19 (suspected or confirmed) was mentioned anywhere on the death certificate from all data sources. The data is from date of death registration and thus there may be a delay between death occurrence and death registration. 
Weeks run from Monday to Sunday and are based on the ISO8601 international standard for week numbering. 
Data are correct as of the week ending the 28$^{th}$ of February 2021.

\subsection{Statistical Agency Data}
\label{sec:data}

%\subsection{Office for National Statistics (ONS)}
The Office for National Statistics (ONS) records weekly provisional counts of the number of deaths registered in England and Wales \cite{ONS}. These data include a break-down of England and Wales for total deaths and COVID-19 related deaths 
that occur in Homes, Hospitals (acute or community, not psychiatric), Hospices, Care Homes, Other Communal Establishment (OCE) (e.g. schools for people with learning disabilities; holiday homes and hotels; common lodging houses; aged persons’ accommodation; assessment centres; schools; convents and monasteries; nurses’ homes), or Elsewhere. Note that this data includes all deaths where COVID-19 is mentioned on the death certificate and is not limited to deaths within 28-days of a positive COVID-19 test. Data from the ONS provides these breakdowns for England and Wales separately and collectively. 

%\subsection{National Records of Scotland (NRS)}
The National Records of Scotland (NRS) \cite{NRS} records the total and COVID-19 related deaths that occur in Care Homes, Homes / Non-institutions, Hospitals and Other institutions (e.g. clinics, medical centres, prisons and schools) recorded at the health board and council area level. We sum the health board deaths for each place of occurrence for both COVID-19 and total death rates to compare to the ONS data. We treat Other institutions from the NRS data as equivalent to OCE in the ONS data given their definitions. 

The Northern Ireland Statistics and Research Agency (NISRA) \cite{NISRA} only records data on the place of occurrence for COVID-19 deaths at the weekly level and the total number of deaths from all causes in these places is not available. The total number of deaths and deaths due to COVID-19 for place of occurrence at the monthly level from January to June 2020 were obtained following an ad-hoc data request to NISRA. We use the monthly data for all calculations involving the total number of deaths at each location and use the weekly data for calculations involving only COVID-19 deaths.  

We combine data from the ONS, NRS and NISRA to get data for the entire UK. We use the combined totals for England and Wales published by ONS as these include non-residents \cite{ONS}. The total deaths in Northern Ireland at the weekly level is not available and omitted in the UK total deaths calculations. Data for deaths due to COVID-19 for Northern Ireland is included in our calculations of COVID-19 deaths in the UK. The monthly NISRA data show there have been 9,014 total deaths and 830 COVID-19 deaths in Northern Ireland between January and June 2020. This compares to 208,989 total deaths and 51,072 COVID-19 deaths from the rest of the UK over the same period, hence the omission of the Northern Ireland data from the total deaths in the UK is likely to have a negligible impact.   

The death data, $d$, processed such that is in the form of a $p\times n$ matrix,
\begin{equation}
d = 
     \begin{pmatrix}
   d_{11} & \dots & d_{1p} \\
   d_{12} & \dots & d_{2p} \\
   \dots & \ddots & \dots \\
   d_{n2} & \dots & d_{np} \\
     \end{pmatrix}
\end{equation}
where $p$ is the number of place of occurrence locations, and $n$ is the number of time points in the dataset. Here $p$=6 representing the six locations in the data (Homes, Hospitals, Hospices, Care Homes, OCE and Elsewhere) and $n$=28 to reflect the number of time points for each location in the data. The time points are weekly counts for total deaths $d^T$ or deaths due to COVID-19 $d^C$ across the $p$ locations. Clearly $d$ is a function of time.

\begin{figure}[t]
         \centering
         \includegraphics[width=0.45\textwidth,trim = 10em 25em 9em 28em, clip=true]{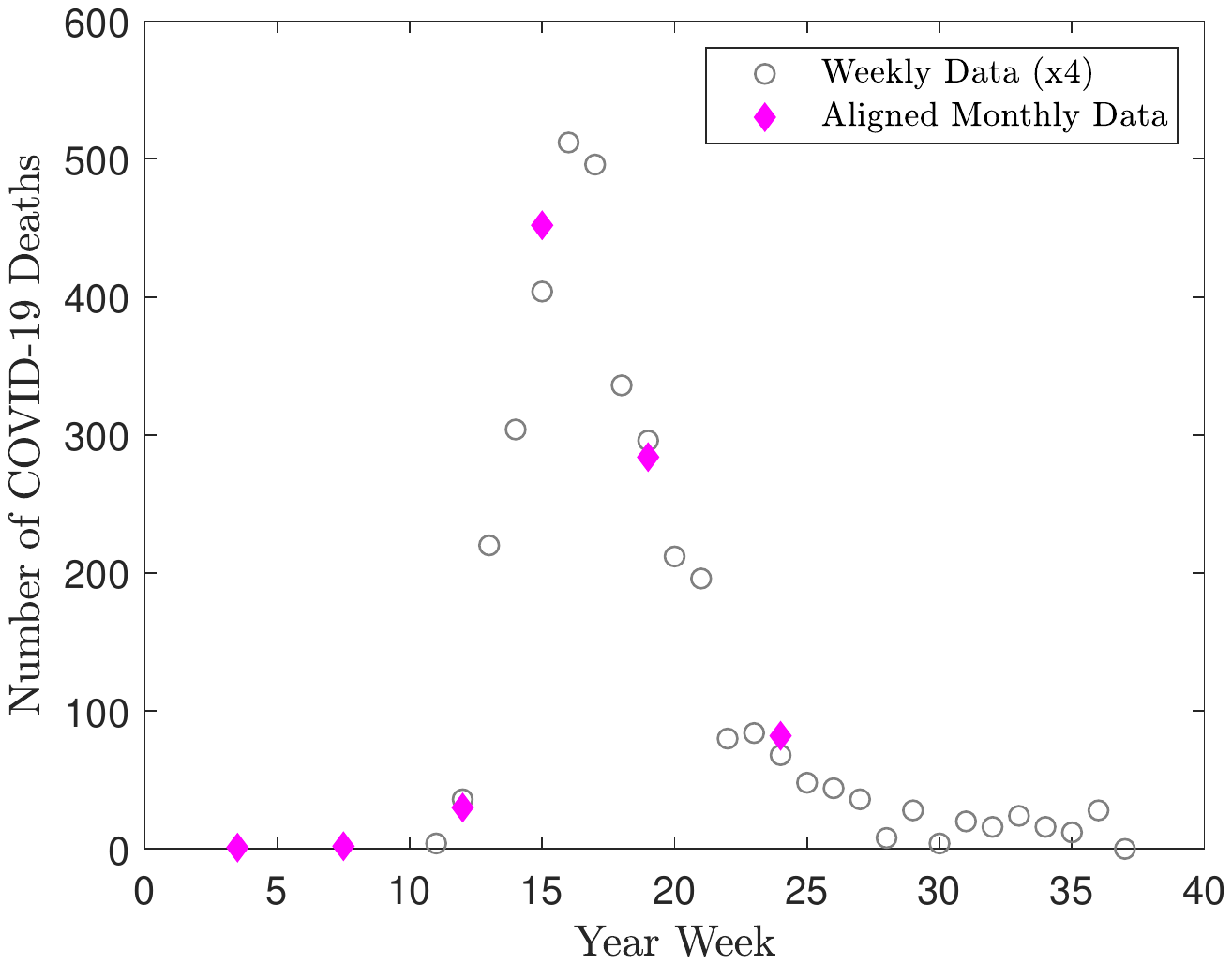}
         \caption{{Aligning monthly and weekly COVID-19 death data in Northern Ireland.} The weekly data is scaled for comparison with the magnitude of the monthly data. All aligned points remain within their original month. Note that data is only available for the first wave of the pandemic.}
         \label{fig:NIalign}
\end{figure}

\subsection{Aligning Northern Ireland Data}
\label{sec:alignNI}
The Northern Ireland Statistics and Research Agency (NISRA) \cite{NISRA} only records data on the place of occurrence for COVID-19 deaths at the weekly level with the equivalent data for deaths from all causes not available. Following an ad-hoc data request to the NISRA place of occurrence data for deaths due to COVID-19 and all causes were provided at monthly intervals between Jan and June 2020. The monthly data is used to calculate the proportion of deaths due to COVID-19 in Northern Ireland, defined in Eq.~(\ref{eq:totalCOVID}). The weekly data is used to calculate the proportion of COVID-19 deaths occurring at each place of occurrence, defined in Eq.~(\ref{eq:COVIDnorm}), due to its better time resolution. 

For comparison to data from the rest of the UK, we align the Northern Ireland COVID-19 deaths from all places in the monthly data to the COVID-19 deaths from all places in the weekly data, see Fig.~\ref{fig:NIalign}. The year week assigned to each point remains within the month the point originated from, that is no points have been moved to an different month based on this alignment. The aligned data is used for comparison with the rest of the UK in terms of peak timing and magnitude.

\subsection{Deaths due to COVID-19}
\label{sec:dataAnalysis}
 
We propose the use of several dynamic normalisations for the place of occurrence death data in order to compare different time points, locations and population across the UK. These also account for the differences in population across the UK and for temporal changes to total number due to the disease evolution. %, i.e. differences in the $R_0$ value. Not only does this remove the effect of differences in exponential growth rates, it also has the added benefit of removing the dependence on any enforced regulations such as lockdown that will be implemented differently in the UK and affect the places of occurrence differently. 
%We normalise the data as follows. 
%Firstly, we look at the percentage of COVID-19 deaths of all deaths, $\mathcal{D}^{CT}$, for each region in the UK. 
We normalise the number of COVID-19 deaths $d^C$ to the total number of deaths $d^T$ (including COVID-19) for a given week for each place of occurrence $i$, 
\begin{equation}
\label{eq:locationCOVID}
\mathcal{D}_i = \frac{d^C_i}{d^T_i}~.
\end{equation} 
$\mathcal{D}$ is a $p$ by $n$ matrix as $d$. Summing $\mathcal{D}_i$ yields the proportion of all deaths that are due to COVID-19 in each region of the UK, formally, 
\begin{equation}
\label{eq:totalCOVID}
{\mathcal{D}^{T} } =  \sum_i \mathcal{D}_i ~.
\end{equation} 
When normalised as Eq.~(\ref{eq:locationCOVID}), and equivalently  Eq.~(\ref{eq:totalCOVID}), the data follow a Weibull type distribution \cite{weibull}. Similarly, infection transmission of COVID-19 also follows a Weibull distribution \cite{ferretti_quantifying_2020, grassly_comparison_2020}. We also modify the model from a traditional Weibull distribution by replacing the typical coefficient $\alpha$ $\beta^{-1}$ with $\gamma$, which is proportional to $ \alpha$ $\beta^{-1}$, as this leads a higher R$^2$ when modelling the data. The resulting model is a modified Weibull distribution given as 
\begin{equation}
\label{eq:Weibull}
    W(t) = \gamma \left(\frac{\left(t-\mu\right)}{\alpha}\right)^{-\beta -1} e^{-\left(\frac{\left(t-\mu\right)}{\alpha}\right)^{-\beta}}~,
\end{equation}
for time point $t$ and $\mu$ the earliest point in the weekly data. Here $\alpha$,  $\beta$ and $\gamma$ are fitting parameters that relate to the maximum value of the distribution ($\gamma$), the variable at the peak of the distribution ($\alpha$), and the shape of the distribution ($\beta$). This model follows the distribution of COVID-19 deaths in the data for each wave of the disease by selecting $\mu$. Note We are using the form of the inverse Weibull distribution (also known as Fr\'{e}chet distribution) \cite{de_gusmao_generalized_2011} which includes a $-$ sign before each $\beta$ term in the equation. This is due to the first wave following the inverse form, with latter waves follow the original Weibull form (see Section~\ref{sec:results}). For brevity and clarity we use only one form of the Weibull distribution (the inverse) and quote negative values for $\beta$ in latter waves of the pandemic even though it is a shape parameter.

\subsection{Proportion of COVID-19 Deaths}
We also normalise the total number of COVID-19 deaths at each location $d^C_i$ to the total number of COVID-19 deaths at all locations for a given week. This dynamic normalisation yields the proportion of COVID-19 deaths at location $i$ for each time step, formally, 
\begin{equation}
\mathcal{D}^C_i =  \frac{d^C_i}{\sum\limits_{i}  d^C_i}~.
\label{eq:COVIDnorm}
\end{equation} 
% not sure whether to include this or not 
% Finally, we normalise the number of  deaths at each location $D^T_i$ to the total number of all deaths at all locations $D^T$ for each week
% \begin{equation}
% \mathcal{D}^T_i = \frac{D^T_i}{ \sum\limits_{i} D^T_i}.
% \end{equation} 
%Similarly the proportion of total deaths at each location, $\mathcal{D}^T_i$, gives an indication of changes in relative deaths (COVID-19 and Excess) between the places of occurrence, and how these distributions change over time.   
%We show that these normalisations given a unique perspective on the data and trends in the different locations. Specifically, they provide a view of the data that shows level of COVID-19 mortality, and its evolution in time, at each location ($\mathcal{D}^{CT}_i$). Furthermore, they show where the highest proportion of COVID-19 deaths are occurring ($\mathcal{D}^C_i$) for a given time point that is irrespective of whether the disease is in an exponential growth phase or declining due to strategies such as lockdown. 
This yields the proportion of COVID-19 deaths at each place of occurrence. %Similarly to $d$ and $\mathcal{D}$, $\mathcal{D}^{C}$ is also a $p$ by $n$ matrix. 
This normalisation is well described by a double sigmoid function (details can be found in Section~\ref{sec:results}),
\begin{equation}
\label{eq:doubleSigmoid}
    f(t) = \lambda \frac{1}{1+e^{-\nu_g(t-\kappa_g)}}  \left(1 -  \frac{1}{1+e^{-\nu_d(t-\kappa_d)}} \right)~.
\end{equation} 
%where $\lamdba$, $\nu_g$, $\kappa_g$, $\nu_d$ and $\kappa_g$ are fitting parameters. 

Here the first part of the equation describes the growth of the curve and the second part (in parenthesis) describes the decay of the curve with time $t$. Parameter $\lambda$ controls the maximum value in the curve, $\nu_{g}$ and $\nu_{d}$ control the steepness of the growth and decay components respectively. Parameter $\kappa_{g}$ is the mid point of the  growth curve and $\kappa_{d}$ is the mid point of the decay term. 

Other distributions have been considered in place of Eq.~(\ref{eq:Weibull}) and Eq.~(\ref{eq:doubleSigmoid}), including log-normal and L\'{e}vy distributions and polynomials of several orders, however, these did not produce satisfactory fits to the data in terms of R$^2$ and thus have been omitted. 

\begin{figure}
    \centering
    \includestandalone[width=.45\textwidth]{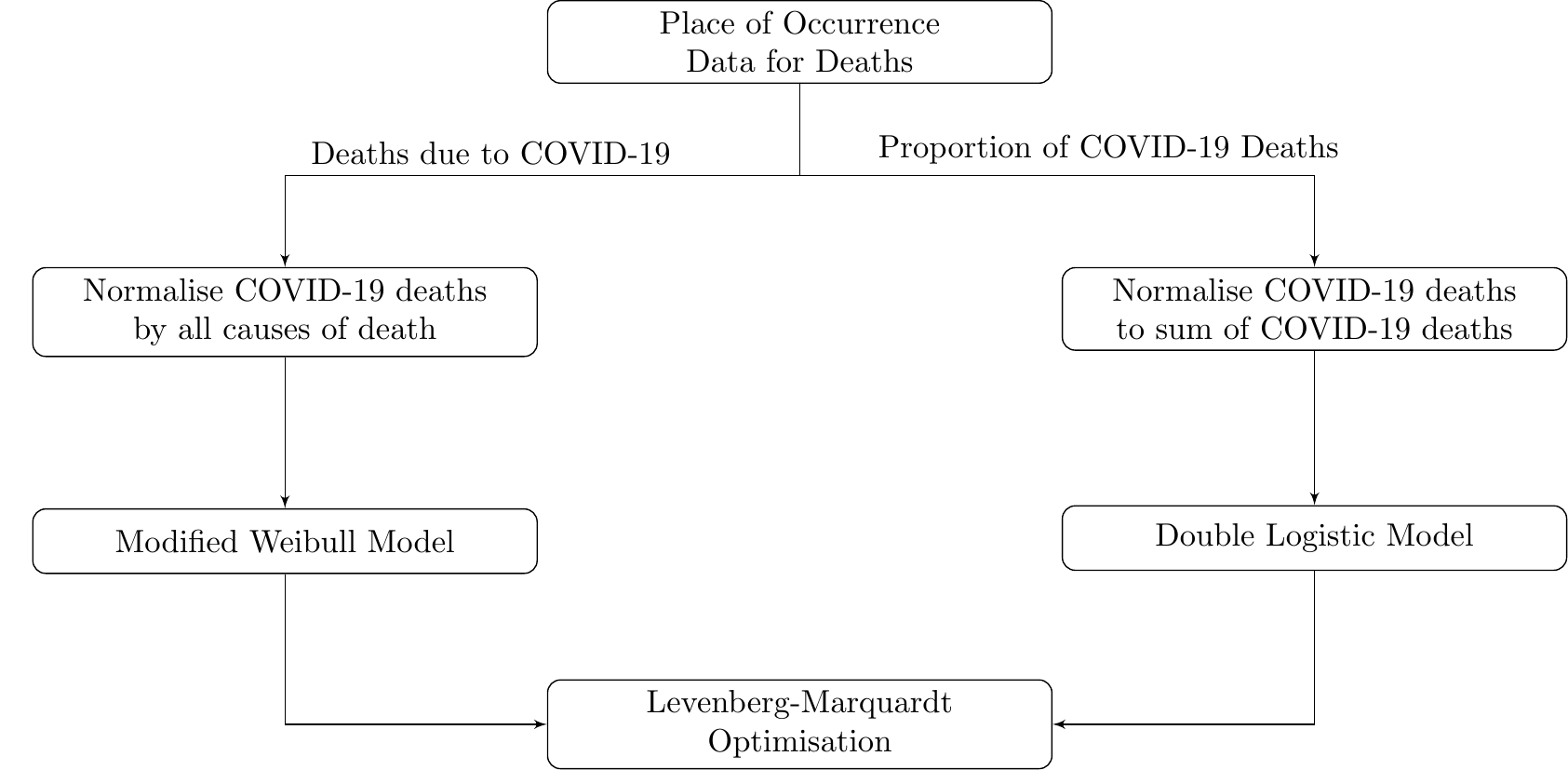}
    \caption{Workflow for modelling the statistical agency data on COVID-19 deaths at the place of occurrence level}
    \label{fig:workflow}
\end{figure}

\subsection{Levenberg-Marquardt Optimisation}
For both models in Eq.~(\ref{eq:Weibull}) and Eq.~(\ref{eq:doubleSigmoid}) using Levenberg-Marquardt Optimisation in order to fit the data. For the optimisation of parameters $\theta$ in the model $\vec{g}(\theta)$, 
\begin{equation}
    \label{eq:optimise}
    %\hat{\beta} \in \textit{argmin}_\beta S(\beta) \equiv \textit{argmin}_\beta \sum_{i=1}^{m} \left[ y_i - f(x_i,\beta) \right]^2
    \hat{\theta} \in \textit{argmin}_\theta ~ S(\theta) \equiv \textit{argmin}_\theta ||\vec{\mathcal{D}} - \vec{g}(\theta)||^2 ~,
\end{equation}
where $\vec{g}(\theta)$ is either Eq.~(\ref{eq:Weibull}) or Eq.~(\ref{eq:doubleSigmoid}), we use the Levenberg-Marquardt nonlinear least squares \cite{Seber2003}. This iteratively updates the estimate of $\theta$ with $\theta + \delta$ where $ \vec{g}(\theta + \delta)$ is approximated as
\begin{equation}
    %f(x_i, \beta + \delta) \approx f(x_i, \beta) + \frac{\partial f (x_i, \beta)}{\partial \beta}\delta~.
    \vec{g}(\theta + \delta) \approx \vec{g}(\theta) + \vec{J}\delta~,  
\end{equation}
where $\vec{J}$ is the Jacobian matrix. Combining this with Eq.~(\ref{eq:optimise}), we obtain 
$ S(\theta) \approx ||\vec{\mathcal{D}} - \vec{g}(\theta) - \vec{J}\delta||^2$, and setting the derivative with respect to $\delta$ to zero yields the Newton Method for optimising $\theta$ 
\begin{equation}
    \left( \vec{J^TJ}\right ) \delta = \vec{J^T}\left[\vec{\mathcal{D}}-\vec{g}\left(\theta\right) \right] ~.   
\end{equation}
The Levenberg-Marquardt method extends this with the inclusion of a non-negative damping factor $\omega$ to increase flexibility and robustness of the optimisation \cite{Levenberg1944}.
\begin{equation}
    \left( \vec{J^TJ} + \omega\vec{I}\right ) \delta = \vec{J^T}\left[\vec{\mathcal{D}}-\vec{g}\left(\theta\right) \right] ~,   
\end{equation}
where $\vec{I}$ is the identify matrix. We fit all curves in this work using the Levenberg-Marquardt method due to its ability to solve nonlinear equations and ill-conditioned problems \cite{ahookhosh2019}. The optimisation is run with a stopping criteria of a minimum tolerance of 10$^{-4} \vec{g}$ at each step and a maximum of 200 iterations. 
%Full details are given in appendix~\ref{sec:dataAnalysis}. We used MATLAB (R2019a) version 9.6.0.1072779 for all analysis. 
The full workflow for the data analysis is given in Fig.~\ref{fig:workflow}.

\section{Results and Discussions}
\label{sec:results}

\subsection{Deaths due to COVID-19 National Level}
\label{sec:resultstotal}
\begin{figure}[t]
\centering
  \includegraphics[width=0.45\textwidth,trim = 10em 25em 9em 28em, clip=true]{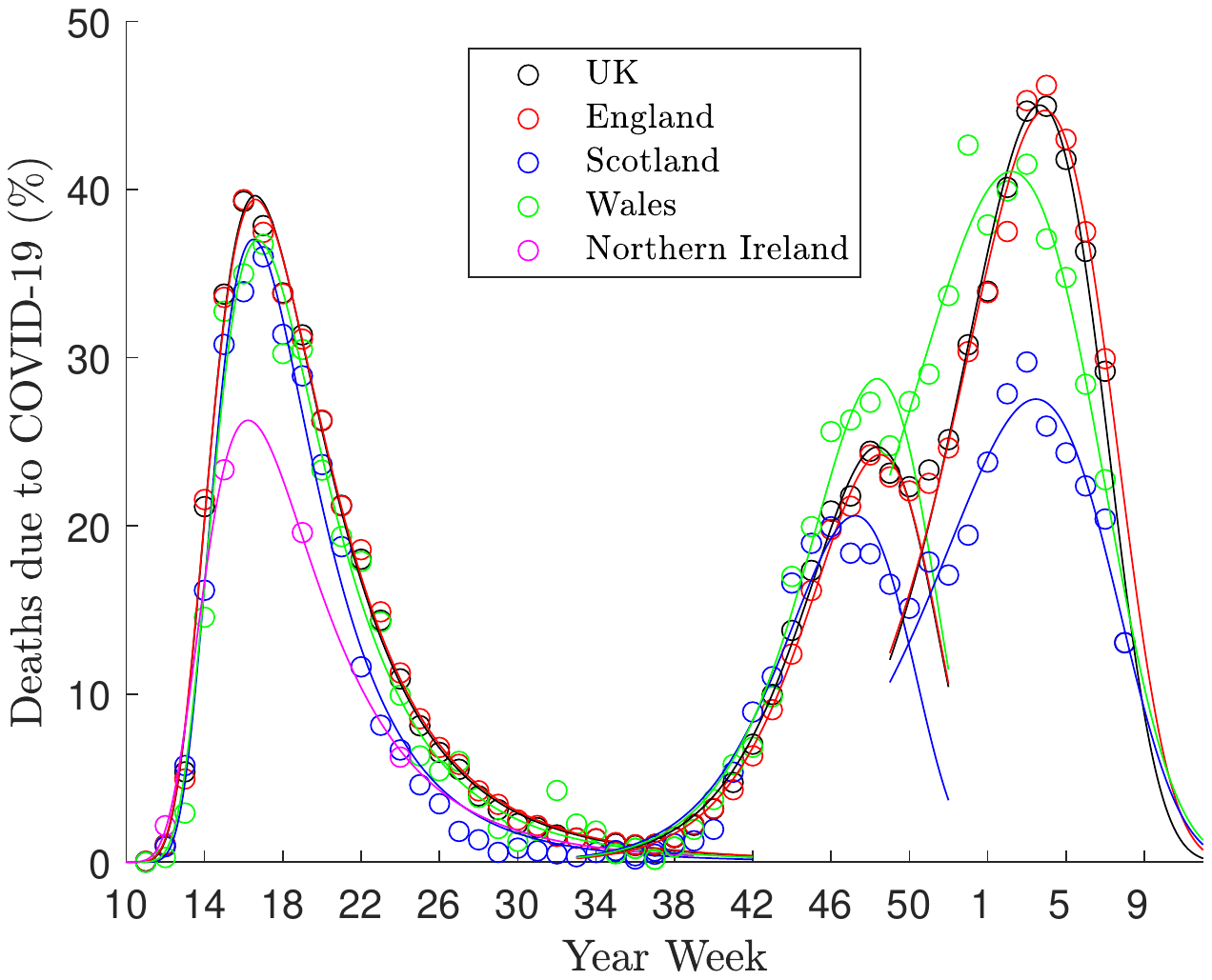}  
  \caption{{Death due to COVID-19, $\mathcal{D}^{T}$, in England, Scotland, Wales, Northern Ireland and the UK.} Curves are modified Weibull distributions fitted to the data. %Peaks occur between between 13 April and April 19, 2020 for all nations. 
  The Northern Ireland data is monthly and has been aligned with the COVID-19 deaths from the weekly Northern Ireland data (see section \ref{sec:alignNI}, Fig.~\ref{fig:NIalign}) and only available for the first wave.}
  \label{fig:DeathsCOVIDoveerlay}
\end{figure}

The proportion of deaths due to COVID-19 in all nations in the UK are well described by the modified Weibull distribution (see Fig.~\ref{fig:DeathsCOVIDoveerlay}). Each of the three waves of the pandemic are well described (R$^2\geq$ 0.95) and the overall distribution is a combination of the individual models for each wave. It is possible to extend the model to optimise the number of waves to the data, however this would increase the number of parameters and we do not have sufficient data for this (28 time points). Hence we fit each wave separately with the first, second and third wave fitted between weeks 10 and 38, 38 and 51 in 2020, and 51 and 8 ranging 2020 and 2021, respectively. 

These models allow us to obtain descriptors of the pandemic at each wave and compare these across the nations in the UK. England closely follows the distribution for the entire UK as it accounts for the majority of the population. Notably for the first wave, deaths due to COVID-19 peak simultaneously across the UK at $\approx$40\%, except Northern Ireland which peaks earlier and reaches a lower maximum. There are clear differences in the second and third waves in the UK. All second wave peaks are below the first wave. Scotland peaks one week earlier than the rest of the UK, and with maximum 4.7\% lower than England and 8.9\% lower than Wales. The third wave also demonstrates large differences across the UK, peaking first in Wales, followed by Scotland and then England 1 and 1.5 weeks later respectively. Only Scotland had a peak in the third wave that was below the first wave, 10.7\% lower, with England's third peak 5.3\% higher than the first wave and Wales increasing by 4.1\%. We note that the data for Northern Ireland is only available for the first waves. 

\subsection{Deaths due to COVID-19 Place of Occurrence Level}
\label{sec:resultsPlace}
%here 

\begin{figure*}
     \centering
     \subfloat[UK\label{fig:COVIDlocationUK}]{%
         \centering
         \includegraphics[width=0.45\textwidth,trim = 10em 26em 9em 27.5em, clip=true]{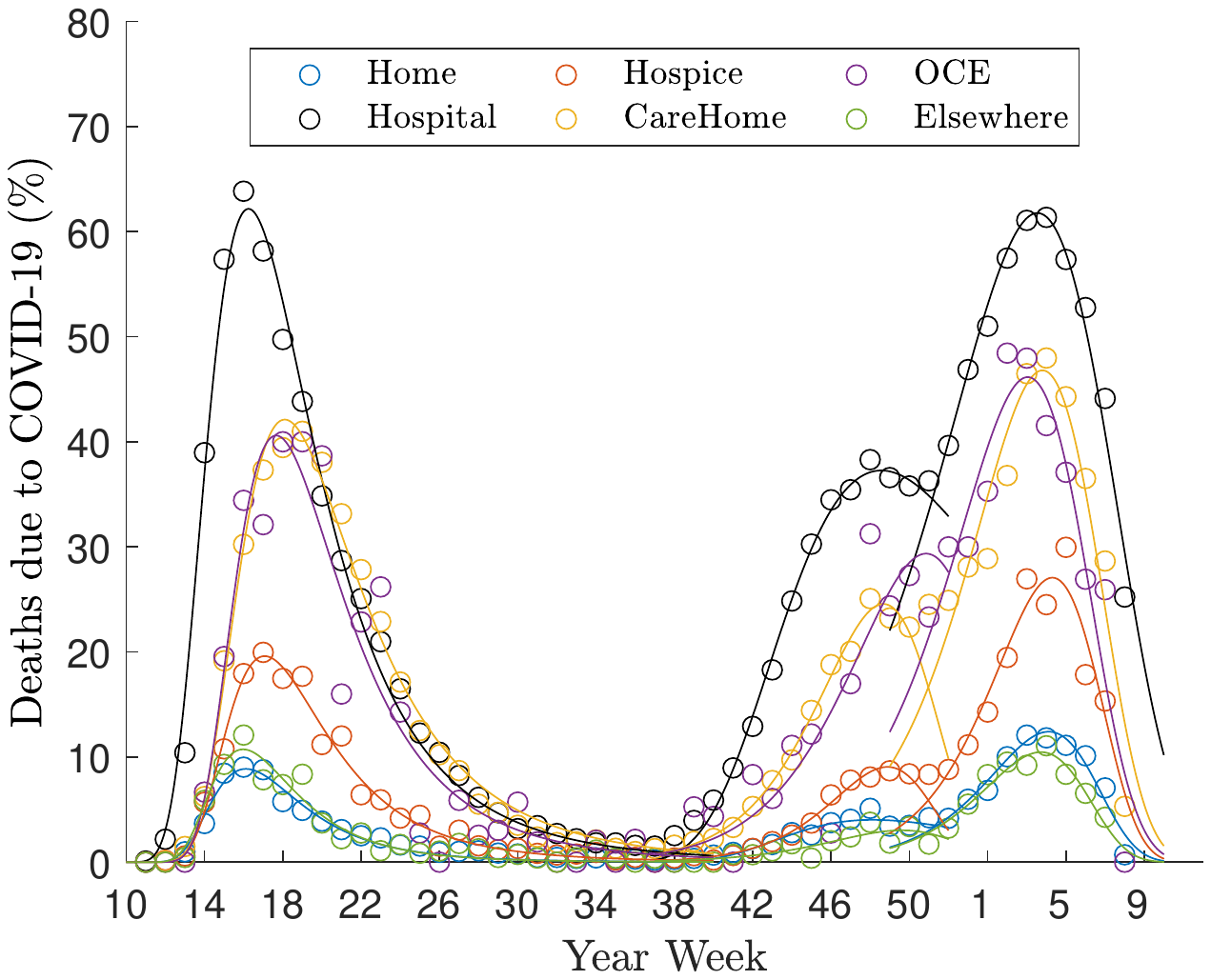}}
     \hfill
     \subfloat[England\label{fig:COVIDlocationENGLAND}]{%
         \centering
         \includegraphics[width=0.45\textwidth,trim = 10em 26em 9em 27.5em, clip=true]{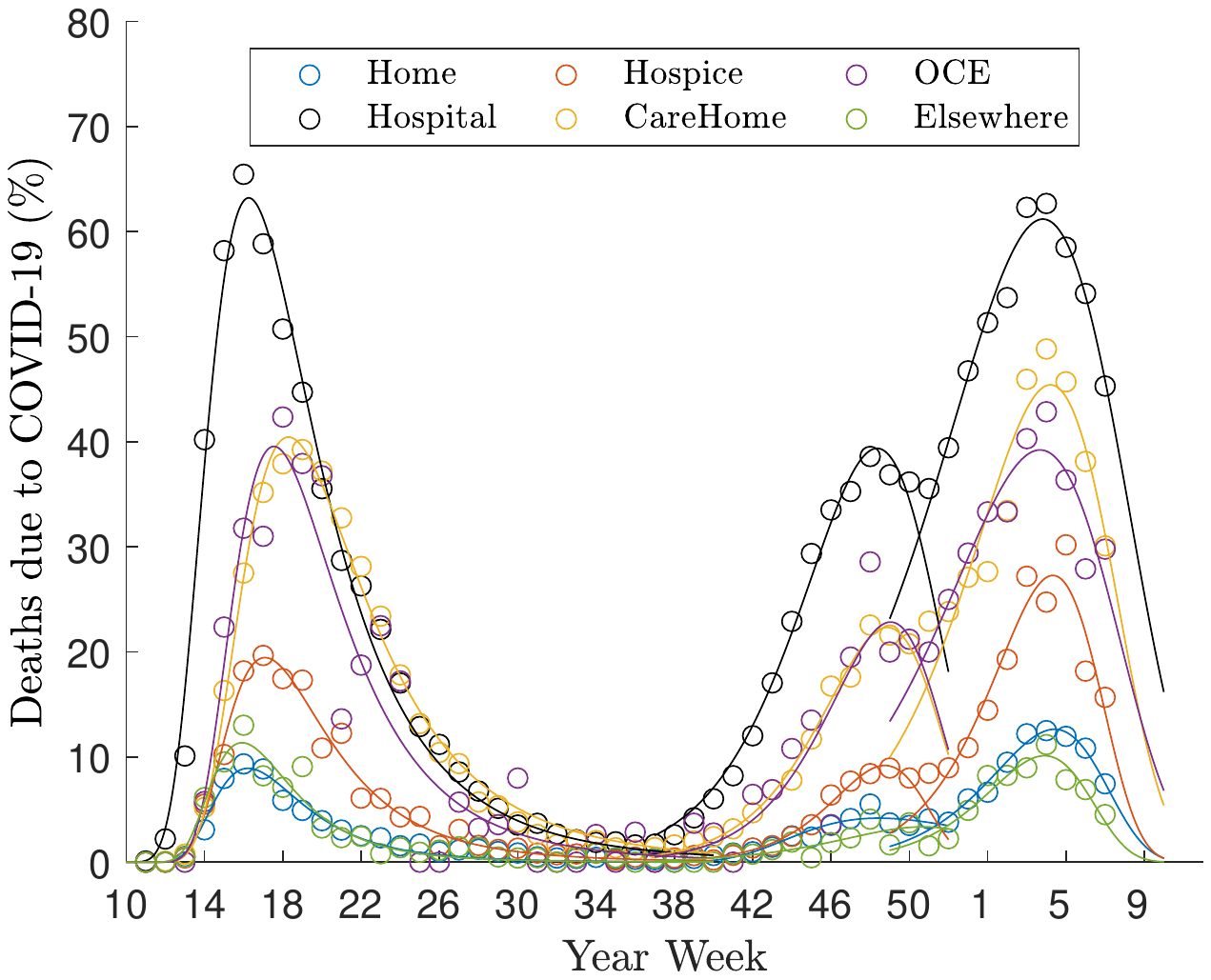}} 
     \\ % \hfill
     \subfloat[Scotland\label{fig:COVIDlocationSCOTLAND}]{%
         \centering
         \includegraphics[width=0.45\textwidth,trim = 10em 26em 9em 27.5em, clip=true]{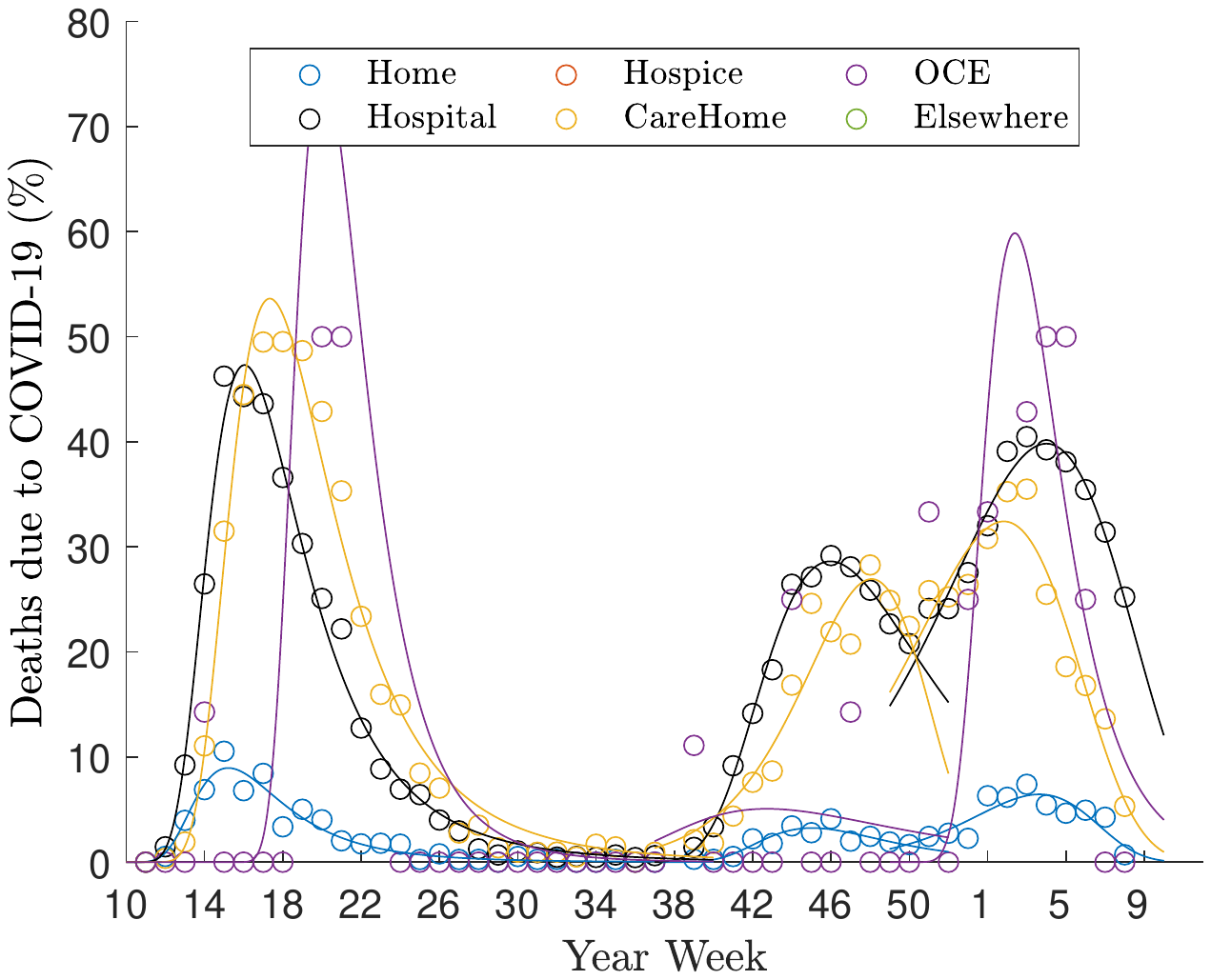}}
     \hfill
     \subfloat[Wales\label{fig:COVIDlocationWALES}]{%
         \centering
         \includegraphics[width=0.45\textwidth,trim = 10em 26em 9em 27.5em, clip=true]{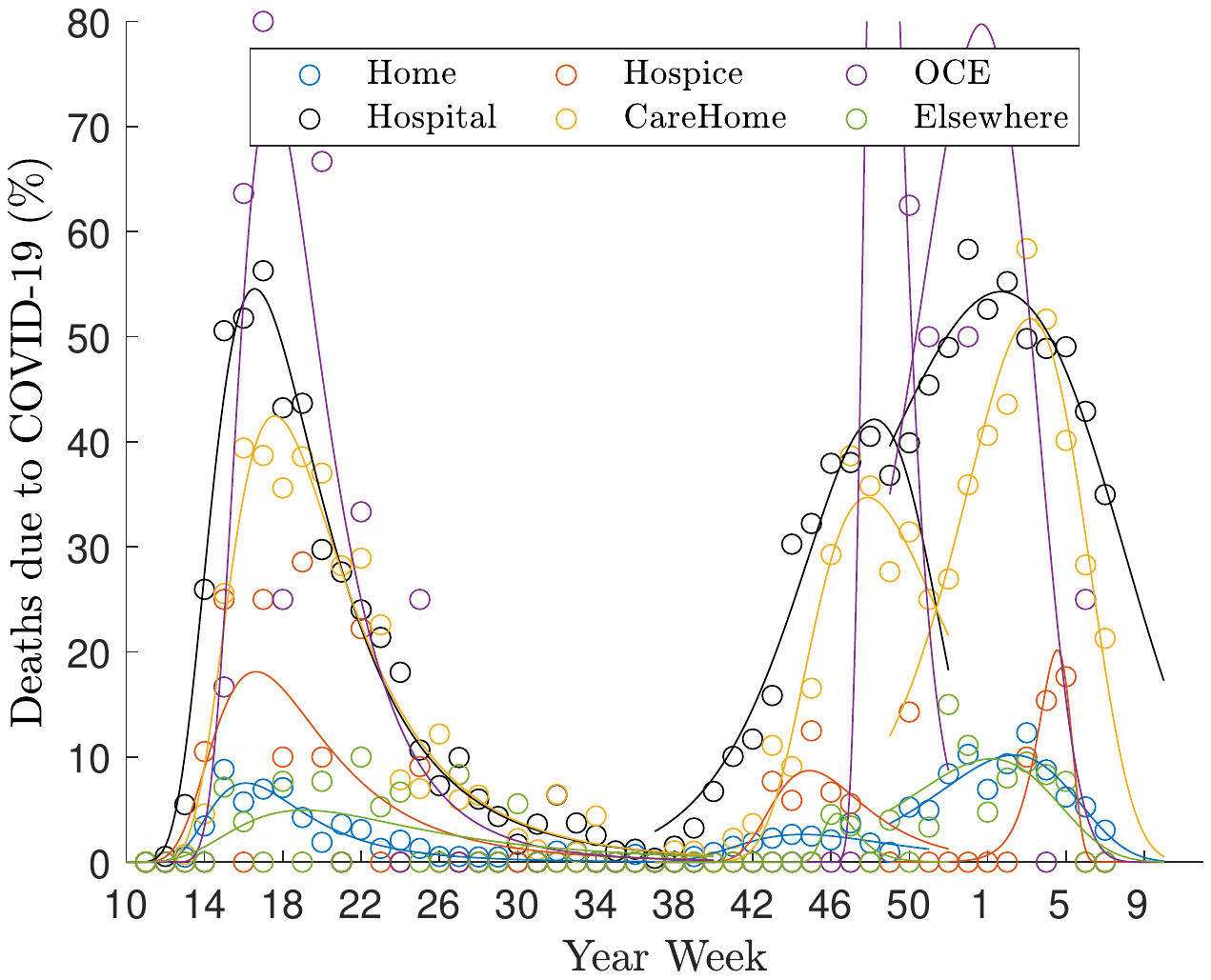}}
    \\
        \caption{{Proportion of deaths due to COVID-19 at place of occurrence across the UK, $\mathcal{D}$, fitted to a modified Weibull distribution.}
        Note that no data for Hospice or Elsewhere are available for Scotland.
        The fluctuations in Hospices, OCEs and Elsewhere are due to small sample sizes. }
        \label{fig:COVIDlocation}
\end{figure*}

\begin{figure}
     \subfloat[Northern Ireland $\mathcal{D}$ \label{fig:COVIDlocationNI}]{%
          \centering
          \includegraphics[width=0.45\textwidth,trim = 10em 30em 9em 27.5em, clip=true]{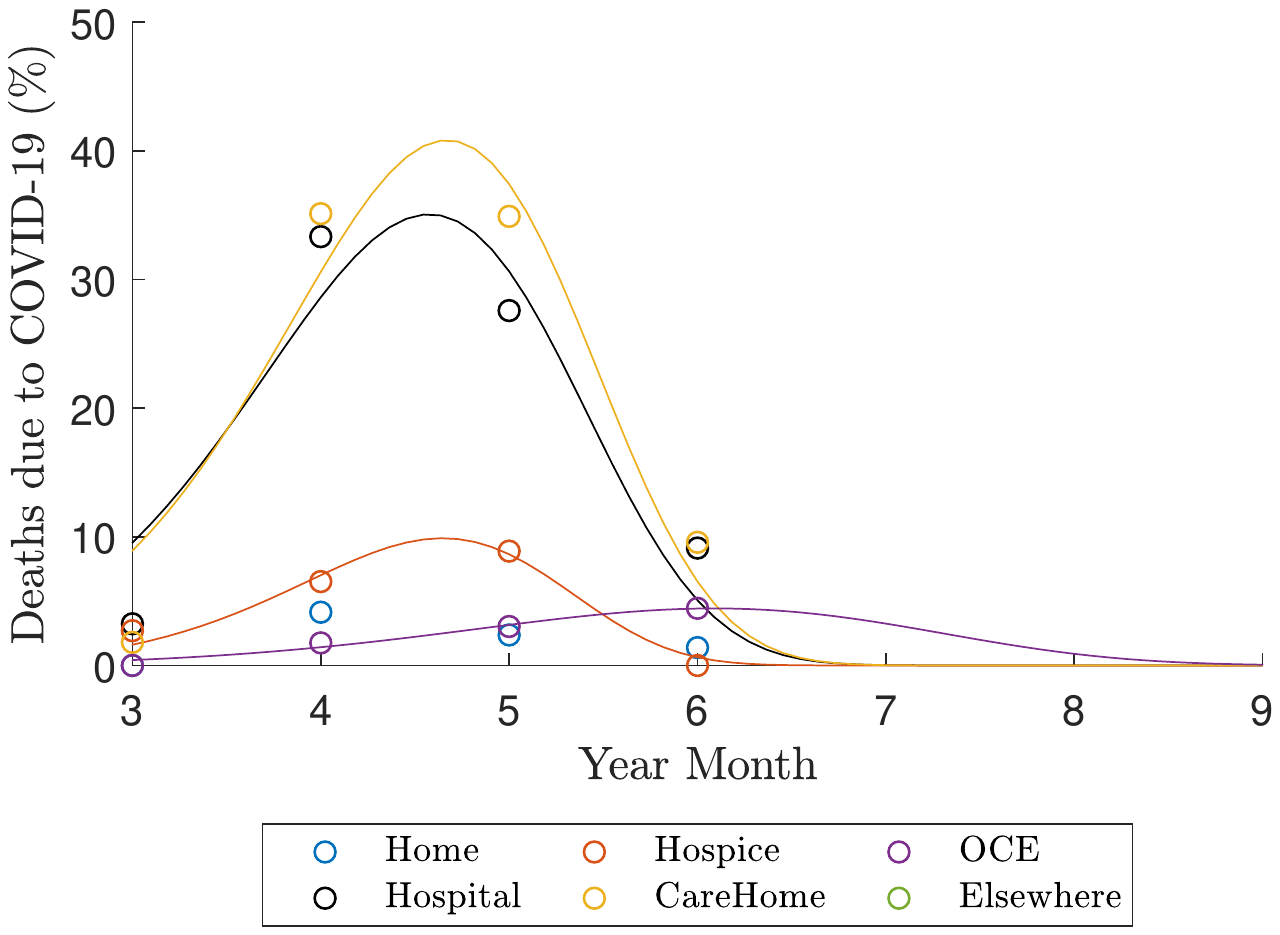}}
    \\
    \subfloat[Northern Ireland $\mathcal{D}^C$ \label{fig:COVIDPropNI}]{%
         \centering
         \includegraphics[width=0.45\textwidth,trim = 10em 30em 9em 27.5em, clip=true]{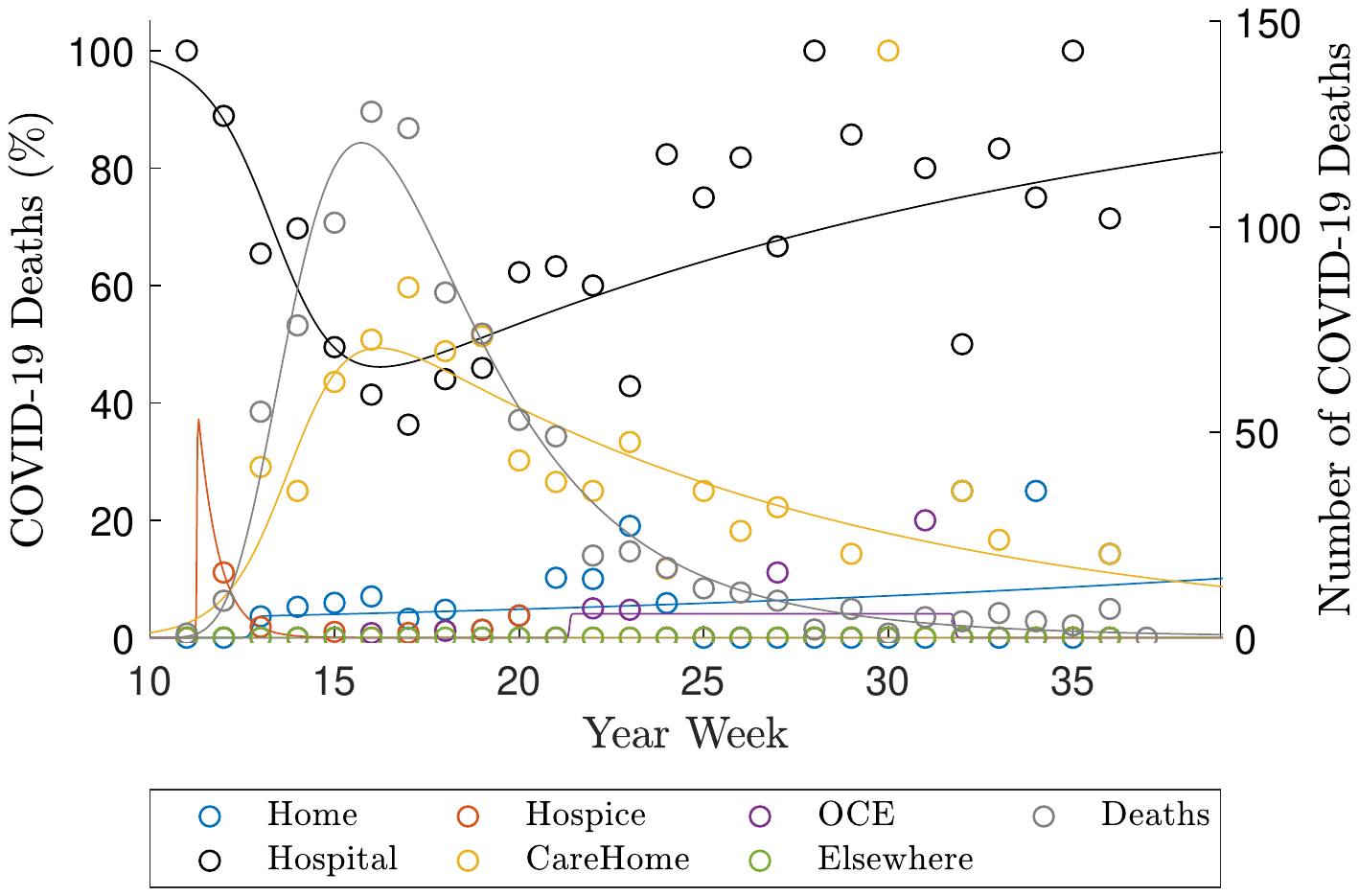}}
        \caption{Northern Ireland data for (a) the deaths due to COVID-19  and (b) proportional of COVID-19 deaths at each place of occurrence. Note only data for the firs waves is available for Northern Ireland. }
        \label{fig:COVIDdeathsNI}
\end{figure}

Deaths due to COVID-19 at the place of occurrence level for all the UK nation is also well described by a modified Weibull distribution (see Fig.~\ref{fig:COVIDlocation}). Across all settings and nations in the UK, the models have $R^2\geq$ 0.80 when more than 100 COVID-19 deaths were recorded at the peak in a given wave. 
%Homes in Scotland and Wales have lower R^2; Scotland >.67 max 63 deaths; wales >.54 max 33 deaths   
Hospice, OCE and Elsewhere settings have lower $R^2$ values for the fits, though this is due to smaller samples sizes with maximums of less than 10 in Wales and Scotland. In England there where a maximum of 140 COVID-19 deaths in Hospice's ($R^2\geq$ 0.87), and a maximum of 40 COVID-19 deaths in both OCE ($R^2\geq$ 0.70) and Elsewhere ($R^2\geq$ 0.61) settings. %All $R^2$ are listed in Table~\ref{tab:Rsquared}. 
Despite the extremely limited number of points in the Northern Ireland data, the modified Weibull distribution models the data well in general, see Fig.~\ref{fig:COVIDlocationNI}. It is worth noting that there are no recorded Elsewhere deaths in the Monthly data from Northern Ireland, and the data For Homes has only three non zero values that appear to be linearly decreasing hence a our model was not fitted to these data. 
We will focus the discussion to those settings with large sample numbers, specifically Homes, Hospitals, Hospices (UK and England), and Care Homes. 

As the trends in England closely follow the UK as a whole, we will limit the discussion to England, Wales, Scotland and Northern Ireland (first wave only) here. The peak in deaths due to COVID-19 in the third wave of the pandemic equalled or exceeded the first wave in all settings in England and Wales, with Scotland experiencing a third wave with fewer deaths due to COVID-19 than in the first wave. For England and Wales, deaths due to COVID-19 reached comparable peaks in the first and third wave in hospitals, but saw significant increases in Care Homes (5\% in England and 9\% in Wales). The deaths due to COVID-19 in Homes also increased by 3\% in both England and Wales, and Hospices in England exhibited an increase of 7\% in the third wave compared to the first. Scotland and Northern Ireland were the only nations to see a higher proportion of deaths due to COVID-19 occurring in Care Homes than Hospitals in the first wave. 

In the first two waves, all nations experienced deaths due to COVID-19 initially peaking in Homes, with peaks in Care Homes and Hospitals occurring one or two weeks later. The only exception to this was the first wave in England, where the peaks in deaths due to COVID-19 in Homes and Hospitals occurred in the same week. Also in England, a peak in Hospices occurred one week after Homes in both the first two waves. This time difference between the peaks is in line with the infectious period of the disease \cite{grassly_comparison_2020} and indicates that the disease may have been transmitted from Homes to Care Homes in the UK and from Homes to Hospices in England.

The third wave exhibited more variation across the UK with England experiencing peaks in Homes, Hospitals and Care Homes within 0.5 weeks, and hospices following shortly after. In Wales, the third waves peaks in deaths due to COVID-19 were first experienced in Hospitals, followed by Homes and Care Homes one and two weeks later respectively. For Scotland, a peak was first encountered in Care Homes followed by Homes and Hospitals two weeks later.

% \begin{table*}[!h]
% \centering
%  \caption{\textbf{Fitting values for inverse Weibull curves.} Ordinary (R$^2_{Ord}$) and adjusted (R$^2_{Adj}$) R$^2$ values, and coefficients for Eq.~(\ref{eq:Weibull}) fits to the proportion of deaths due to COVID-19 for each place of occurrence. Coefficients include standard errors in parentheses. Places that can not be fitted with a Weibull curve are omitted. All R$^2<0.80$ values occur for places with small sample sizes. Any settings with negative $\beta$ value follow a Weibull distribution rather than an inverse Weibull distribution. Note the Northern Ireland data in the top section is based on the aligned data and $\alpha$ corresponds to weeks, where as in the bottom section the Northern Ireland data is based on the unaligned monthly data hence $\alpha$ corresponds to months. }
% \begin{tabular}{l ccccc}
% \toprule 
% Place & R$^2_{Ord}$ & R$^2_{Adj}$ & $\alpha$ & $\beta$ & $\gamma$ \\
% \toprule
% UK  &$1.00 $&$ 1.00 $&$ 17.13 ~(0.05) $&$ 5.33 ~(0.08) $&$ 105.99 ~(1.37) $\\
% England  &$0.99 $&$ 0.99 $&$17.16 ~(0.05) $&$5.26 ~(0.09)$&$105.34 ~(1.55) $\\ 
% Scotland &$ 0.99 $&$ 0.99 $&$16.91 ~(0.07) $&$ 6.04 ~(0.15) $&$ 99.35 ~(2.07) $\\ 
% Wales  &$0.98 $&$ 0.98 $&$17.23 ~(0.09) $&$5.53 ~(0.17) $&$ 99.02 ~(2.54) $\\ 
% Northern Ireland  &$1.00 $&$ 1.00 $&$ 16.80 ~(0.05) $&$ 5.16 ~(0.11) $&$ 70.18 ~(1.02) $\\ 
% \bottomrule \\
%   \label{tab:Rsquared}
%   \end{tabular}
% \end{table*}

\subsection{Implication of $\beta$}
\label{sec:beta}

The sign of $\beta$ in Eq.~(\ref{eq:Weibull}) determines whether the distribution is a (modified) inverted Weibull, $\beta>$ 0, or if it is a (modified) Weibull, $\beta<$ 0. Note the case where $\beta$ = 0, Eq.~(\ref{eq:Weibull}) reduces to $W(t) \propto t^{-1}$. The sign of $\beta$ affects the shape of the distribution, specifically whether the heavy tail is on the right side ($\beta>$ 0) or on the left side ($\beta<$ 0). The implication for COVID-19 is the difference between; a rapid surge in deaths due to COVID-19 with a slow reduction after the peak (positive $\beta$); or a more gradual increase in cases and a rapid recovered after the peak (negative $\beta$). The sign of $\beta$ for each nation and place of occurrence setting is given in Table~\ref{tab:betaSign}.

% NATIONS WIDE 
At the nation level, the first waves are modelled with the inverted Weibull distribution, $\beta>$ 0 (R$^2\geq$ 0.95), however the second and third waves in all nations follow a Weibull distributions, $\beta<$ 0. The latter two waves are well described by the Weibull distribution with R$^2\geq$ 0.95 for all cases, except in Scotland (second and third peak) and Wales (third peak) where R$^2\geq$ 0.86.

%place of occurrence 
At the place of occurrence level, all nations exhibit the same trends for Homes with $\beta>$ 0 for the first and second waves, and $\beta<$ 0 for the third wave. Care Homes and Hospitals had $\beta>$ 0 for the first wave and $\beta<$ 0 for the third waves. Some differences in the UK were experiences in the second wave with $\beta<$ 0 for Hospitals in England and Wales, and for Care Homes in England and Scotland, with $\beta>$ 0 in all other settings.

\begin{table}[t]
\centering
 \caption{Sign for $\beta$ in Eq.~(\ref{eq:Weibull}) for Deaths due to COVID-19 in the UK at the national and place of occurrence levels. Positive signs (+) represent a distribution with a rapid growth and slow recovery (heavy right side), and negative sign (-) represent a slow growth and rapid recovery (heavy left side). NA indicates no data available.}
\begin{tabular}{lccc}
\toprule 
Place & $\beta$ Wave 1 & $\beta$ Wave 2 & $\beta$ Wave 3   \\
\toprule
UK          &  +    & -  &  -  \\
England     & +     &  - &  -  \\
Scotland    &  +    &  - &   - \\
Wales       &  +    &  - &  -  \\
Northern Ireland &  +  & NA  &  NA  \\
\bottomrule 
Homes (UK)          &  +  &  + &  -  \\
Homes (England)     &  +  &  + &  -  \\
Homes (Scotland)    &  +  &  + &  -  \\
Homes (Wales)       &  +  &  + &  -  \\
\bottomrule 
Care Homes (UK)         &  +  &  - &  -  \\
Care Homes (England)    &  +  &  - &  -  \\
Care Homes (Scotland)   &  +  & -  &  -  \\
Care Homes (Wales)      &  +  & +  &  - \\ 
\bottomrule 
Hospitals (UK)      & +   & +  &  -  \\
Hospitals (England) &  +  &  - &  -  \\
Hospitals (Scotland) &  +  &  + &  -  \\
Hospitals (Wales)   &  +  & -  &  -  \\
\bottomrule 
\label{tab:betaSign}
\end{tabular}
\end{table}

\subsection{Proportion of COVID-19 Deaths}
\label{sec:resultsProportion}
When normalising to the number of COVID-19 deaths at each place of occurrence to the total number of COVID-19 deaths for each week, Eq.~(\ref{eq:COVIDnorm}), the data broadly follow a double logistic distribution described by Eq.~(\ref{eq:doubleSigmoid}). The data deviate from this model when the sample size is small between the first and second waves when there are few COVID-19 deaths. There appears to be some finer structure occurring before the second wave and at the peak of the third wave. More data maybe required to capture this additional behavior.

\begin{figure*}
     \centering
  \subfloat[UK\label{fig:COVIDlocationUK}]{%
         \centering
         \includegraphics[width=0.45\textwidth,trim = 10em 26em 9em 27.5em, clip=true]{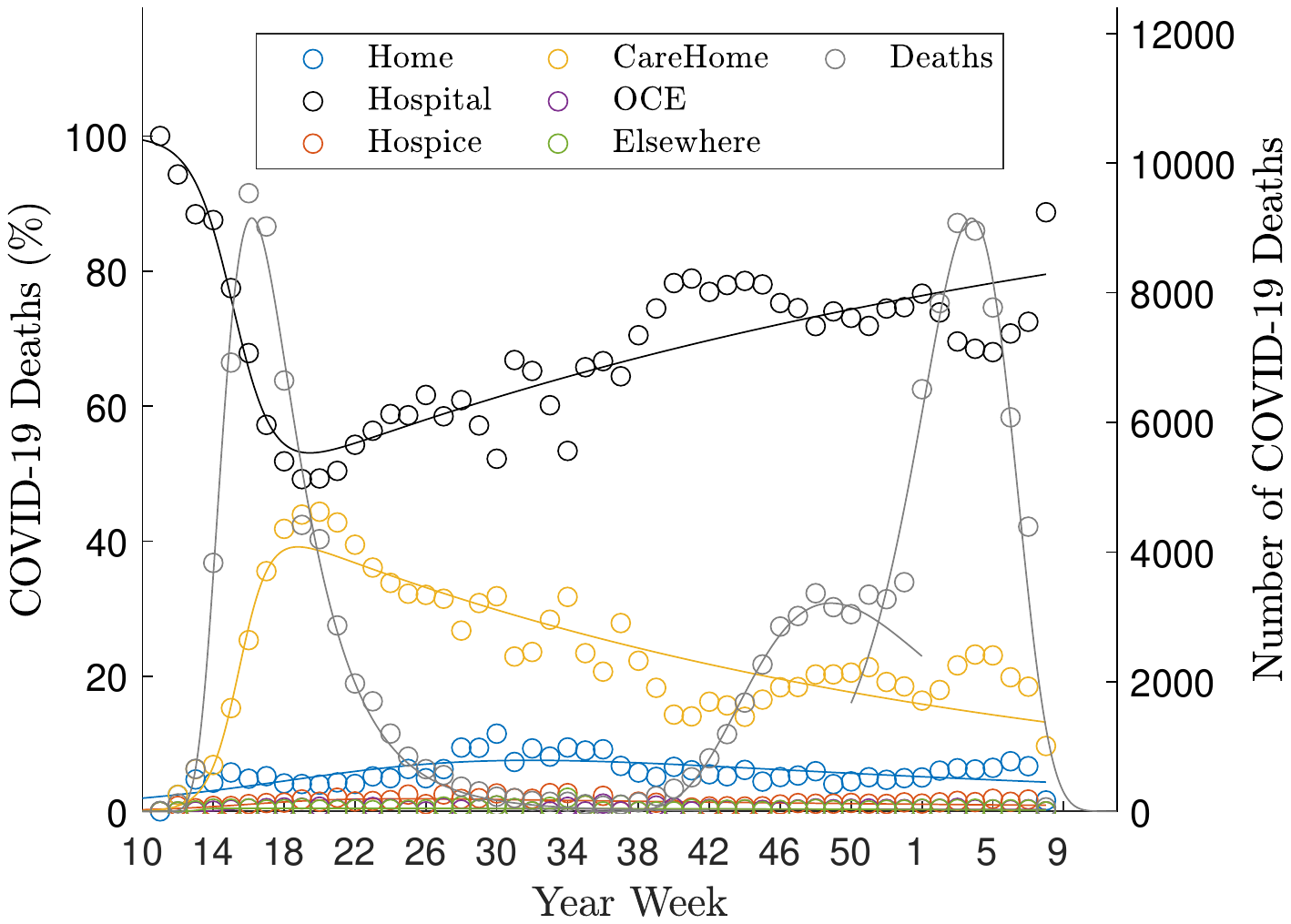}}
     \hfill
  \subfloat[England\label{fig:COVIDlocationENGLAND}]{%
         \centering
         \includegraphics[width=0.45\textwidth,trim = 10em 26em 9em 27.5em, clip=true]{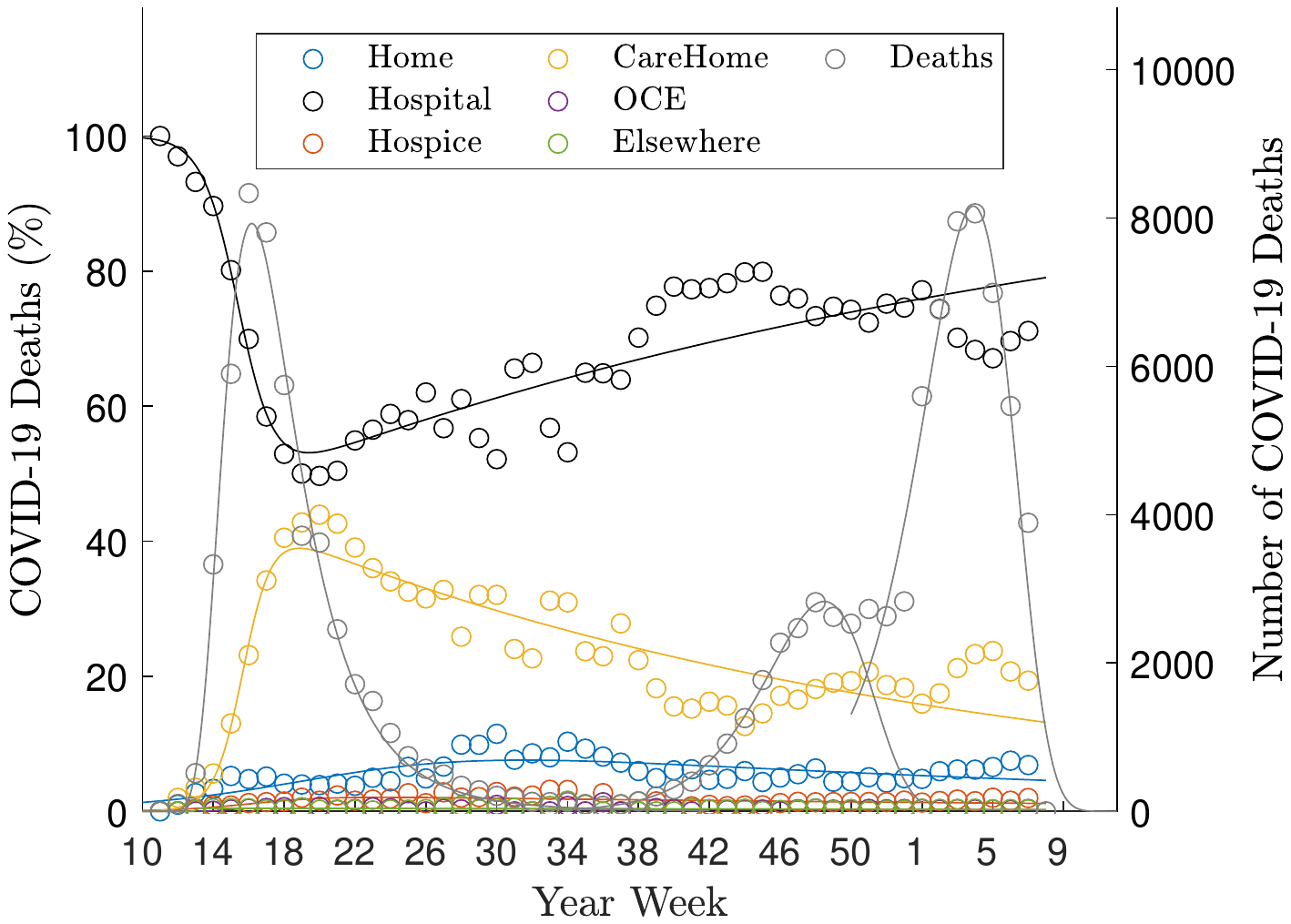}}
     \\
  \subfloat[Scotland\label{fig:COVIDlocationSCOTLAND}]{%
         \centering
         \includegraphics[width=0.45\textwidth,trim = 10em 26em 9em 27.5em, clip=true]{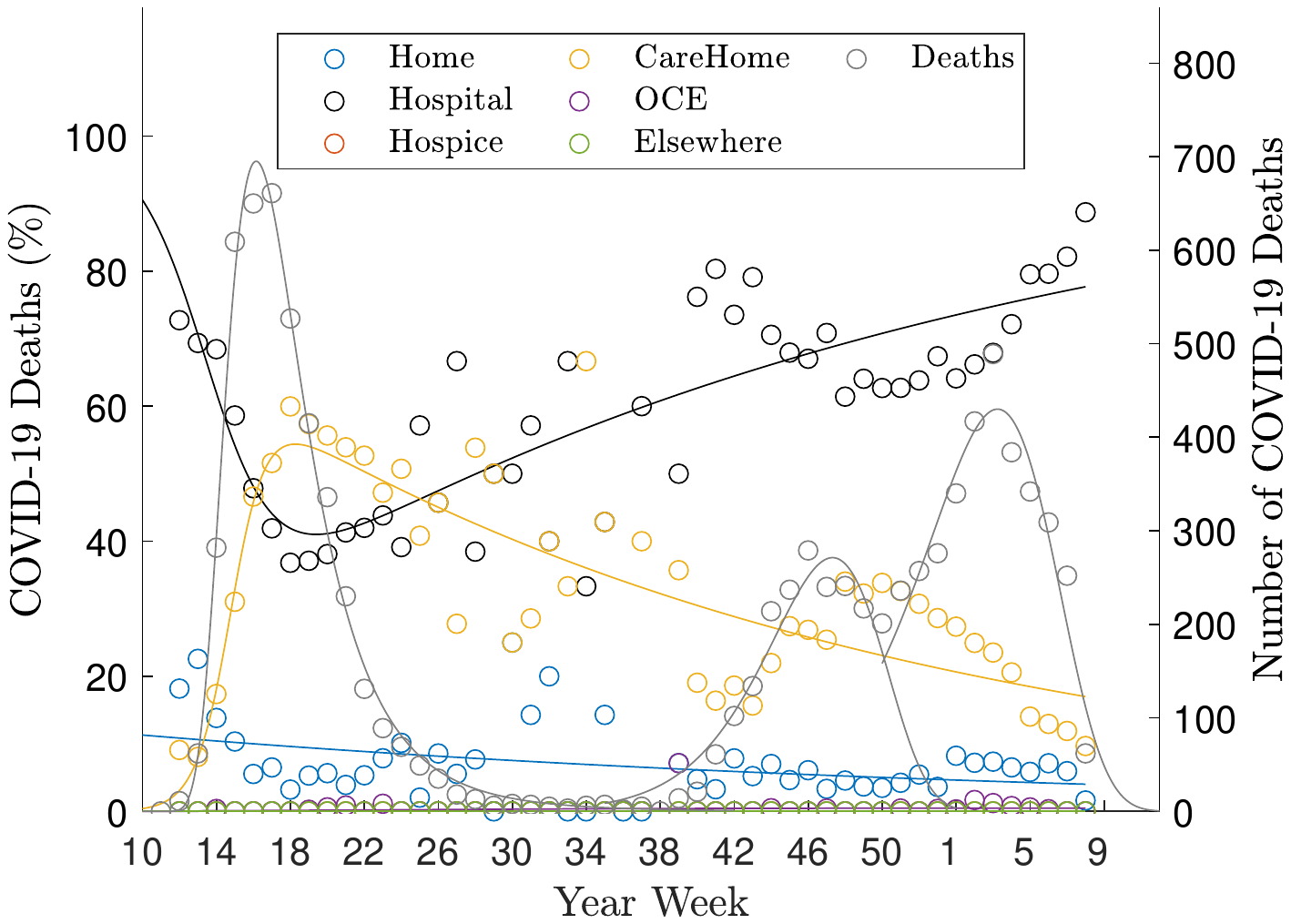}}
    \hfill
  \subfloat[Wales\label{fig:COVIDlocationWALES}]{%
         \centering
         \includegraphics[width=0.45\textwidth,trim = 10em 26em 9em 27.5em, clip=true]{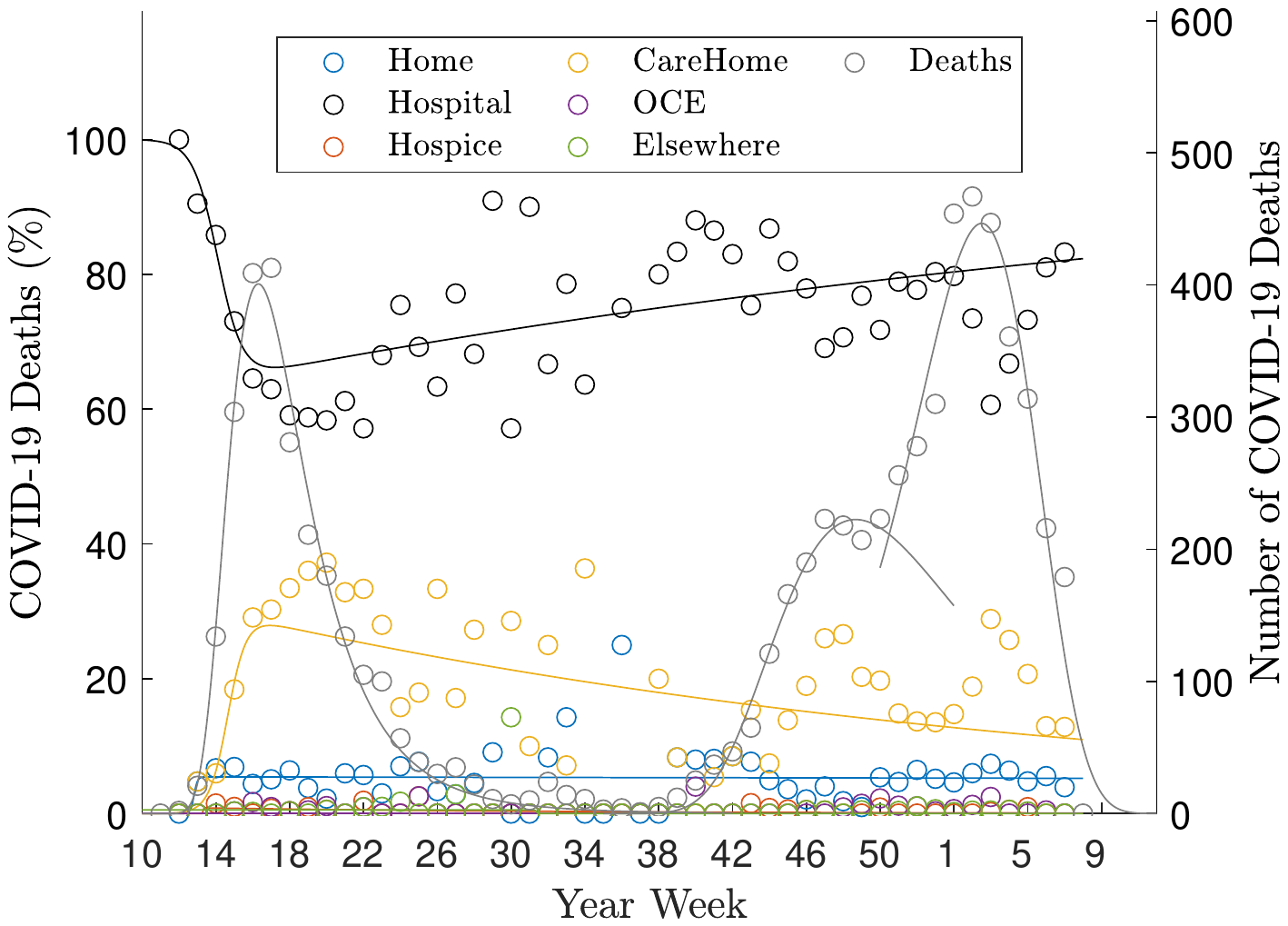}}
    \hfill
        \caption{{Proportion of COVID-19 deaths at place of occurrence across the UK, $\mathcal{D}^C$.} Data are fitted to a double logistic distribution only deviating well past the peak in the number of COVID-19 deaths were sample numbers are small.}
        \label{fig:COVIDdeathsPOL}
\end{figure*}

In the first wave, the greatest contribution to COVID-19 deaths are from Hospitals and Care Homes across nations in the UK. The data for Hospitals, following normalisation as Eq.~(\ref{eq:COVIDnorm}), are fitted to $100 - f(t)$ where $f(t)$ is given in Eq.~(\ref{eq:doubleSigmoid}) as this yielded higher R$^2$ (100 is selected as the scale is percentage). At the start of the pandemic almost all COVID-19 deaths occur in Hospitals. This proportion rapidly decreases reaching a minimum at or after the peak in number of COVID-19 deaths, see Fig.~\ref{fig:COVIDdeathsPOL}. In concordance with a decrease Hospitals, the proportion of COVID-19 deaths that occur in Care Homes rapidly increase at the start of the pandemic reaching a maximum after the peak in number of COVID-19 deaths. The delay between the peaks in total number of COVID-19 deaths and the proportion occurring in Care Homes ($\mathcal{D^C}$) highlights the vulnerability of Care Home residence and need for better protective measures.

After the catastrophic impact in Care Homes during the first wave, the majority of COVID-19 related deaths occurred in Hospitals across the UK, and continued throughout the pandemic. After the reaching a maximum shortly after the peak in total number of COVID-19 deaths in the first wave, the proportion of COVID-19 deaths in Care Homes has continued to decrease gradually. Fluctuations occurring around the second and third waves are not described by our model, which only captures the overall trends here due to insufficient data.
In the second wave, England and Scotland saw the proportion of COVID-19 deaths occurring in Care Homes peak around half the maximum they reached in the first wave. For Wales, the maximum proportion of COVID-19 deaths occurring in Care Homes in the second wave reached approximately 2/3 that of the first wave. In the third wave, England and Wales experience slight increases in the proportion of COVID-19 deaths in Care Homes compared to the second wave, but far below that of the first. Scotland did not experienced another peak in the proportion of COVID-19 deaths in Care Homes following the second wave. Throughout the third wave in Scotland the proportion of COVID-19 deaths in Hospitals increased. In all nations the proportion of COVID-19 deaths in Care Homes is lower in the second and third waves of the pandemic compared to the first, indicating protection for residence may have improved after the first wave. 

The proportion of COVID-19 deaths in Homes reached its largest values between the first and second waves. Although this coincided with easing of lockdown and national restrictions, it is worth noting that this is when the number of COVID-19 deaths is extremely low compared to rest of the pandemic, and is unlikely to correspond to an increased risk. 

\section{Conclusions}

We have presented a modified Weibull distribution that describes the trends in deaths due to COVID-19 across the UK from national statistical agency data. Our model describes the data during each wave of the pandemic at the national and place of occurrence level for each nation in the UK. Additionally, we introduce a double logistic model to broadly describe the proportion of COVID-19 deaths in each setting throughout the pandemic. 

Modeling the data provides insight in to the evolution of the disease mortality across these different settings allowing retrospective analysis of strategic decision making. Due to the coarseness of the available data, models of the data can provide useful descriptors of the impact of COVID-19, such as the ordering of the peaks in different regions and settings. Specifically, we find that in the first two waves of the pandemic, deaths due to COVID-19 first peak in Homes followed by Hospitals and Care Homes typically peaking 1-2 weeks later. This time period is in line with the infectious period of the disease and indicates that transmission to Care Homes, which where catastrophically hit in the first wave, may have originated from Homes. 

Our model includes a shape parameter ($\beta$) whose sign determines whether the distribution has a heavy right or left side, with important implications for COVID-19. Our results show that the first wave is described by a heavy right sided distribution (positive $\beta$) where the deaths due to COVID-19 rapidly increase to a peak then slowly decreasing. The second and third waves on the other hand are described by a heavy left sided distribution (negative $\beta$) where the increase is slower and reduction in deaths is rapid. This difference is potentially critical for national response with a heavy left sided distribution (negative $\beta$) making so called `circuit breaker' lockdowns very effective. Combining this with the slower growth may result in early introduction of circuit breakers being an effective and short term measure for controlling COVID-19 deaths in future waves. These models can also be used to compare the strategic responses across the UK to inform decision making in future waves of the pandemic, of particular importance will be future waves resulting from new variants of the disease. The negative $\beta$ values in the latter waves maybe due to changes in behaviour such as social distancing, masks, hand washing, etc. 

We have analysed place of occurrence death data for COVID-19 and modelled trends across the UK. This analysis can be useful in successive waves of the COVID-19 pandemic, including new strains and potentially in future pandemics. Understanding the evolution of the mortality of the disease is vital for national strategic response and within specific settings such as Care Homes. We discuss the possible implications of the distribution shape (heavy left or heavy right tail) and how this could be useful in strategic response. We have demonstrated the models application to sample sizes differing by two orders of magnitude and suitability to weekly or monthly data. In addition, we model all waves of the pandemic in the UK and place of occurrence settings allowing the determination of ordering of the peaks. This could readily be applied to other nations, or additionally fine grain regional data or other discriminators such as age, gender and ethnicity, and potentially outside of COVID-19 research.

% use section* for acknowledgment
\section*{Acknowledgment}
This work was funded by the UK's Department for Business, Energy \& Industrial Strategy through the national measurement system. We acknowledge the Office for National Statistics (ONS), the National Records of Scotland (NRS) and the Northern Ireland Statistics and Research Agency (NISRA) for collecting and supplying the data. The author would like to thank Louise Wright and Nadia Smith (National Physical Laboratory) for review of the data and manuscript, Matt Hall (National Physical Laboratory) for review of the manuscript, and Elaine Longden (NISRA) for providing the monthly data for Northern Ireland for this work. 

% Can use something like this to put references on a page
% by themselves when using endfloat and the captionsoff option.
\ifCLASSOPTIONcaptionsoff
\newpage
\fi

\bibliographystyle{IEEEtran}  
\bibliography{references}  %%% Remove comment to use the external .bib file (using bibtex).

\end{document}